\documentstyle[preprint,tighten,eqsecnum,aps,prb]{revtex}
\begin{document}
\preprint{cond-mat/9310068}
\title{Random-matrix theory of parametric correlations\\
in the spectra of disordered metals and chaotic billiards}
\author{C. W. J. Beenakker and B. Rejaei}
\address{Instituut-Lorentz, University of Leiden,\\
P.O. Box 9506, 2300 RA Leiden, The Netherlands}
\date{August 1993}
\maketitle
\begin{abstract}
A random-matrix theory is developed for the adiabatic response to an external
perturbation of the energy spectrum of a mesoscopic system. The basic
assumption is that spectral correlations are governed by level repulsion.
Following Dyson, the dependence of the energy levels on the perturbation
parameter is modeled by a Brownian-motion process in a fictitious viscous
fluid. A Fokker-Planck equation for the evolution of the distribution function
is solved to yield the correlation of level densities at different energies and
different parameter values. An approximate solution is obtained by asymptotic
expansion and an exact solution by mapping onto a free-fermion model. A
generalization to multiple parameters is also considered, corresponding to
Brownian motion in a fictitious world with multiple temporal dimensions.
Complete agreement is obtained with microscopic theory.
\end{abstract}
\newpage
\narrowtext

\section{Introduction}
\label{intro}

This is a theoretical investigation of the adiabatic response to an external
perturbation of the energy spectrum of a complex quantum mechanical system. We
consider a Hamiltonian ${\cal H}(X)$ which depends on a parameter $X$. The
$X$-dependence of the energy levels $E_{n}(X)$ shown in Fig.\ 1 by way of
example, is taken from a calculation of the hydrogen atom in a magnetic
field.\cite{Gol91} Only levels with the same cylindrical symmetry are shown. A
weak $X$-dependence of the mean density of states is removed by a rescaling of
the energy.
What remains is an irregular oscillation of $E_{n}$ as a function of $X$. Two
levels which approach each other are repelled as $X$ is increased further,
leading to a sequence of avoided crossings at which the derivative ${\dot
E}_{n}\equiv dE_{n}/dX$ changes sign. The average $\overline{\dot{E}_{n}}$ is
zero, averaged either over a range of $X$ or over a range of $n$. The
correlator of ${\dot E}_{n}(X)$ and ${\dot E}_{m}(X')$ is non-zero for nearby
levels $n,m$ and for nearby parameters $X,X'$, and serves as a quantitative
characterization of how the system responds to an external perturbation.

Our investigation was motivated by a remarkable {\em universality\/} of the
parametric correlations discovered by Szafer and Altshuler.\cite{Sza93} They
considered a disordered metallic particle with the topology of a ring,
enclosing a magnetic flux $\phi$ (measured in units of $h/e$). The energy
levels $E_{i}(\phi)$ depend parametrically on $\phi$. The dispersion is
characterized by the ``current density''
\begin{equation}
j(E,\phi)=\sum_{i}\delta\biglb(E-E_{i}
(\phi)\bigrb)\frac{d}{d\phi}E_{i}(\phi).\label{jdef}
\end{equation}
Szafer and Altshuler applied diagrammatic perturbation theory\cite{Alt86} to
compute the correlation function
\begin{equation}
C(\delta E,\delta\phi)=\overline{j(E,\phi)
j(E+\delta E,\phi+\delta\phi)},\label{Cdef}
\end{equation}
where the overline indicates an average over an ensemble of particles with
different impurity configurations. The result was that the correlator $C(\delta
E,\delta\phi)$ becomes universal for $\delta E=0$,
\begin{equation}
C(0,X)=-\frac{2}{\pi^{2}\beta X^{2}},\label{C0}
\end{equation}
with $\beta=2$ and $X=\delta\phi$.
Eq.\ (\ref{C0}) is universal in the sense that it contains no microscopic
parameters which characterize the particle, such as the diameter $L$, the mean
level spacing $\Delta$, the Fermi velocity $v_{\rm F}$, or the mean free path
$l$. It holds for $(\Delta/E_{\rm c})^{1/2}\ll\delta\phi\ll 1$, where $E_{\rm
c}\simeq\hbar v_{\rm F}l/L^{2}$ is the Thouless energy.

Eq.\ (\ref{C0}) was proven for the case that the randomness in the energy
spectrum is due to scattering by randomly located impurities. Numerical
simulations indicated that it applies generically to chaotic systems, even if
there is no disorder and all randomness comes from scattering at irregularly
shaped boundaries.\cite{Sza93}
(The average in that case is taken over $E$ and $\phi$.) Further work on
disordered systems by Simons and Altshuler,\cite{Sim93} based on a
non-perturbative ``supersymmetry'' formalism,\cite{Efe83} has shown that Eq.\
(\ref{C0}) with $\beta=1$ and $X=\delta U$ applies if the external perturbation
is a spatially fluctuating electrostatic potential $Us({\bf r})$.
(The function $s({\bf r})$ should vary smoothly on the scale of the electron
wavelength, with vanishing spatial average.) These analytical investigations
assumed non-interacting electrons. Recent numerical simulations of a Hubbard
model by Faas, Simons, and Altshuler\cite{Faa93} have shown that Eq.\
(\ref{C0}) remains valid in the presence of electron-electron interactions.
The correlator (\ref{C0}) thus provides a universal quantum mechanical
characterization of the response of a chaotic system to an external magnetic or
electric field. Such universality calls for a {\em random-matrix theory\/} of
parametric correlations. It is the purpose of this paper to present such a
theory.

The basic principle of random-matrix theory (RMT) is that the spectral
correlations are dominated by level repulsion.\cite{Wig65} Level repulsion is a
direct consequence of the Jacobian $\prod_{i<j}|E_{i}-E_{j}|^{\beta}$
associated with the transformation from the space of $N\times N$ Hermitian
matrices ${\cal H}$ to the smaller space of $N$ eigenvalues $E_{i}$. Level
repulsion is universal in the sense that it is fully determined by the symmetry
class of the Hamiltonian ensemble.
There exist just three symmetry classes,\cite{Dys62a} characterized by the
number $\beta=1,2,4$ of independent components of the matrix elements of ${\cal
H}$: $\beta=1$ in zero magnetic field (real ${\cal H}$), $\beta=2$ in non-zero
field (complex ${\cal H}$), and $\beta=4$ for strong spin-orbit scattering in
zero magnetic field (quaternion ${\cal H}$). The three ensembles are called
orthogonal ($\beta=1$), unitary ($\beta=2$), and symplectic ($\beta=4$).

The Wigner-Dyson theory of random matrices yields a level-density correlation
function $K(\delta E)$ which is {\em universal\/} for level separations $\delta
E$ greater than the mean level spacing $\Delta$.\cite{Meh91} The function
$K(\delta E)$ measures correlations between the level density
\begin{equation}
n(E,X)=\sum_{i=1}^{N}\delta\biglb(E-E_{i}(X)\bigrb)\label{rhoXdef}
\end{equation}
at different energies $E$ and $E+\delta E$, but at the {\em same\/} value of
the external parameter $X$:
\begin{equation}
K(\delta E)=\bar{n}(E,X)\bar{n}(E+\delta E,X)-
\overline{n(E,X)n(E+\delta E,X)}.\label{Kdef}
\end{equation}
The universal limiting form of $K$ in the Wigner-Dyson theory is
\begin{equation}
K(\delta E)=\frac{1}{\pi^{2}\beta\delta E^{2}}.\label{K0}
\end{equation}
The universal correlator (\ref{K0}) was first obtained from RMT in the context
of nuclear physics, and then applied to small metallic particles by Gorkov and
Eliashberg.\cite{Gor65} Much later, it was derived from a microscopic
Hamiltonian by Efetov\cite{Efe83} and by Altshuler and
Shklovski\u{\i}.\cite{Alt86}
The microscopic theory shows that Eq.\ (\ref{K0}) holds for a disordered metal
in the energy range $\Delta\ll\delta E\ll E_{\rm c}$. Numerical simulations
have established that the Wigner-Dyson theory applies generically to systems
with chaotic classical orbits,\cite{Chaos} and also that it remains valid in
the presence of electron-electron interactions.\cite{Mon93}

The level-density correlation function (\ref{K0}) is thus universal in the same
sense as the parametric correlation function (\ref{C0}). This suggests that it
should be possible to derive Eq.\ (\ref{C0}) by some extension of the
Wigner-Dyson theory to parameter-dependent Hamiltonians ${\cal H}(X)$.
We will show that the {\em Brownian-motion model\/} used by Dyson\cite{Dys62}
to construct a parameter-dependent ensemble of random matrices, yields
parametric correlations in agreement with the microscopic theory of Altshuler,
Simons, and Szafer.\cite{Sza93,Sim93}.

The outline of this paper is as follows. In Sec.\ II we formulate the problem
of a random-matrix theory of parametric correlations and define the mapping
onto Dyson's Brownian-motion model. The correlation functions which we will
calculate are summarized in Sec.\ III. In Sec.\ IV we present an asymptotic
analysis which yields the correlation functions in the limit that the dimension
$N$ of the Hamiltonian matrix goes to infinity.
An exact result for the correlation functions in the Brownian-motion model for
a special ensemble is given in Sec.\ V, and compared with the large-$N$ result
of the previous section. In Sec.\ VI we extend the theory to parametric
correlations involving multiple parameters. We conclude in Sec.\ VII, by
comparing the results of random-matrix theory with the microscopic theory.

The results of the asymptotic analysis were briefly announced in a recent
Letter.\cite{Bee93b}

\section{Brownian-motion model}
\label{brownian}

Starting point of our analysis is Dyson's Brownian-motion model\cite{Dys62} for
the evolution of an ensemble of $N\times N$ Hermitian matrices as a function of
an external parameter $\tau$.
Dyson's idea was to regard $\tau$ as a fictitious ``time'', and to model the
$\tau$-dependence of the distribution of eigenvalues $P(\{E_{n}\},\tau)$ by the
one-dimensional Brownian motion of $N$ classical particles at positions
$E_{1}(\tau),E_{2}(\tau),\ldots E_{N}(\tau)$, in a fictitious viscous fluid
with friction coefficient $\gamma$ and temperature $\beta^{-1}$. Level
repulsion is accounted for by the interaction potential $-\ln|E-E'|$ between
particles at $E$ and $E'$.
The particles move in a confining potential $V(E)$, which is determined by the
density of states.

With these definitions, $P(\{E_{n}\},\tau )$ evolves according to the
Fokker-Planck equation\cite{Dys62}
\begin{eqnarray}
&&\gamma\frac{\partial P}{\partial\tau}=
\sum_{i=1}^{N}\frac{\partial}{\partial E_{i}}
\left(P\frac{\partial W}{\partial E_{i}}+\beta^{-1}
\frac{\partial P}{\partial E_{i}}\right),\label{FokkerPlanck}\\
&&W(\{E_{n}\})=-\sum_{i<j}\ln|E_{i}-E_{j}|+\sum_{i}V(E_{i}).\label{Wdef}
\end{eqnarray}
Eq.\ (\ref{FokkerPlanck}) has the $\tau\rightarrow\infty$ (``equilibrium'')
solution
\begin{equation}
P_{\rm eq}(\{E_{n}\})=Z^{-1}{\rm e}^{-\beta W},\label{Peq}
\end{equation}
where $Z$ is such that $P_{\rm eq}$ is normalized to unity. Eq.\ (\ref{Peq}),
for $\beta=1$, $2$, and $4$, is the eigenvalue distribution in the orthogonal,
unitary, and symplectic ensemble.\cite{Meh91} It has the form of a Gibbs
distribution, with the symmetry index $\beta$ playing the role of inverse
temperature. The fictitious energy $W$ contains a logarithmic repulsive
interaction plus a confining potential $V$.
The function $V(E)$ is chosen such that $P_{\rm eq}$ yields the required
average eigenvalue density (which depends on microscopic parameters, but is
assumed to be independent of $\tau$). The logarithmic interaction has a
fundamental geometric origin: The factor $\exp(\beta\sum_{i<j}\ln|E_{i}-E_{j}|)
=\prod_{i<j} |E_{i}-E_{j}|^{\beta}$ is the Jacobian associated with the
transformation from the space of Hermitian matrices ${\cal H}$ to the smaller
space of eigenvalues $E_{n}$.

The $N$-dimensional Fokker-Planck equation (\ref{FokkerPlanck}) is equivalent
to $N$ coupled Langevin equations,
\begin{equation}
\gamma\frac{dE_{i}}{d\tau}=-\frac{\partial W}
{\partial E_{i}}+{\cal F}_{i}(\tau),\; i=1,2,\ldots N.\label{Langevin}
\end{equation}
The random force ${\cal F}$ is a Gaussian white noise of zero mean,
$\overline{{\cal F}_{i}(\tau)}=0$, and variance
\begin{equation}
\overline{{\cal F}_{i}(\tau){\cal F}_{j}(\tau')}=
\frac{2\gamma}{\beta}\delta_{ij}\delta(\tau-\tau').\label{varF}
\end{equation}
The Fokker-Planck equation (\ref{FokkerPlanck}) and the Langevin equations
(\ref{Langevin}) are equivalent levels of description of the Brownian
motion.\cite{Kam81}

The fictitious time $\tau$ needs still to be related to the perturbation
parameter $X$ in the Hamiltonian ${\cal H}(X)$ of the physical system one is
modeling. Furthermore, we need a microscopic interpretation of the coefficient
$\gamma$. These issues were not addressed in Ref.\ \onlinecite{Dys62}, but are
crucial for our purpose. Let $\tau=0$ coincide with $X=0$, so that
\begin{equation}
P(\{E_{n}\},0)=\prod_{i=1}^{N}
\delta(E_{i}^{\vphantom{0}}-E_{i}^{0}),\label{Pinitial}
\end{equation}
with $E_{i}^{0}$ the eigenvalues of ${\cal H}(0)$. For $\tau>0$ we then
identify
\begin{equation}
\tau=X^{2}.\label{tauX}
\end{equation}
This is the simplest relation between $\tau$ and $X$ which is consistent with
the average initial rate of change of the energy levels: On the one hand,
\begin{equation}
\overline{(E_{i}(X)-E_{i}^{0})^{2}}=X^{2}
\overline{\left(\frac{dE_{i}}{dX}\right)^{2}}+{\cal O}(X^{3})
\end{equation}
is of order $X^{2}$ for small $X$, while on the other hand the ensemble average
\begin{equation}
\overline{(E_{i}(\tau)-E_{i}^{0})^{2}}=
\frac{2\tau}{\beta\gamma}+{\cal O}(\tau^{2})
\end{equation}
is of order $\tau$ for small $\tau$, according to Eqs.\ (\ref{Langevin}) and
(\ref{varF}). The identification (\ref{tauX}) also implies the relation
\begin{equation}
\frac{2}{\beta\gamma}=\overline{\left(
\frac{dE_{i}}{dX}\right)^{2}}\label{gammarelation}
\end{equation}
between the friction coefficient and the mean-square rate of change of the
energy levels.

Eq.\ (\ref{FokkerPlanck}) or (\ref{Langevin}) is the simplest description of
the Brownian motion of the energy levels which is consistent with the
equilibrium distribution (\ref{Peq}). It is not the most general description:
1. One could include the ``velocities'' $dE_{n}/d\tau$ as independent
stochastic variables, and work with a $2N$-dimensional evolution equation. In
the case of Brownian motion in a physical fluid, the appropriate evolution
equation is Kramer's equation.\cite{Kam81}
It describes the dynamics of a Brownian particle on the time scale of the
collisions with the fluid molecules. Since the viscous fluid in Dyson's
Brownian-motion model is fictitious, it is not clear what the appropriate
$2N$-dimensional evolution equation should be in this case.  2. One could let
$\gamma$ be a matrix function $\gamma_{ij}(\{E_{n}\})$ of the configuration of
energy levels. Such a configuration dependence (known in fluids as hydrodynamic
interaction) would be an additional source of correlations, which is ignored.
That is the basic assumption of Dyson's Brownian-motion model, that the
spectral correlations are dominated by the fundamental geometric effect of
level repulsion.
The Brownian-motion model is known to provide a rigorous description of the
transition between random-matrix ensembles of different symmetry.\cite{Len90}
However, there exists no derivation of Eq.\ (\ref{FokkerPlanck}) or
(\ref{Langevin}) from a microscopic Hamiltonian. Here we apply the
Brownian-motion model to fluctuations around equilibrium in the random-matrix
ensembles (\ref{Peq}), and show that there is a complete agreement with the
microscopic theory for disordered metals.\cite{Sza93,Sim93}

\section{Correlation functions}
\label{correlation}

We consider observables $A(X)$ of the form
\begin{equation}
A(X)=\sum_{i=1}^{N}a\biglb( E_{i}(X)\bigrb).\label{Adef}
\end{equation}
A quantity of the form (\ref{Adef}) is called a {\em linear statistic\/} on the
eigenvalues of ${\cal H}(X)$. The word ``linear'' indicates that $A$ does not
contain products of different eigenvalues, but the function $a(E)$ may well
depend non-linearly on $E$. We assume that $a$ varies smoothly on the scale of
the mean level spacing $\Delta$. [In particular, this excludes the case of a
step function $a(E)$.]
The correlator of $A$ at two parameter values $X$ and $X'$ is $\overline{\delta
A(X)\delta A(X')}$, where $\delta A\equiv A-\bar{A}$. The overline denotes an
average over a range of $X$ at constant $\delta X\equiv X'-X$ (or,
alternatively, over an ensemble of statistically equivalent systems). Of
particular interest is the integrated correlator
\begin{equation}
\chi_{_A}=\int_{0}^{\infty}\!d\,\delta X\,
\overline{\delta A(X)\delta A(X+\delta X)}.\label{chidef}
\end{equation}
To compute the correlator of an arbitrary linear statistic we need the density
correlation function
\begin{eqnarray}
S(E,X,E',X')&=&\sum_{i,j}\overline{\delta\biglb(E-E_{i}(X)\bigrb)
\delta\biglb(E'\!-E_{j}(X')\bigrb)}\nonumber\\
&&\mbox{}-\left(\sum_{i}\overline{\delta\biglb(E-E_{i}(X)\bigrb)}
\right)\left(\sum_{j}\overline{\delta\biglb(E'-E_{j}(X')\bigrb)}
\right).\label{SXdef}
\end{eqnarray}
The correlator of $A$ at $X$ and $X'$ then follows from a double integration,
\begin{equation}
\overline{\delta A(X)\delta A(X')}=
\int_{-\infty}^{\infty}\!dE\int_{-\infty}^{\infty}
\!dE'\,a(E)a(E')S(E,X,E',X').\label{ASXdef}
\end{equation}

We will also consider the correlator $\overline{\dot{A}(X)\dot{A}(X')}$ of the
derivative $\dot{A}\equiv dA/dX$ of the linear statistic (\ref{Adef}). This
correlator follows from the density correlation function $S$ by
\begin{equation}
\overline{\dot{A}(X)\dot{A}(X')}=
\int_{-\infty}^{\infty}\! dE\int_{-\infty}^{\infty}\!
dE'\,a(E)a(E')\frac{\partial^{2}}
{\partial X\partial X'}S(E,X,E',X').\label{AdotSXdef}
\end{equation}
Alternatively, we can compute the correlator of $\dot{A}$ from the current
correlation function
\begin{equation}
C(E,X,E',X')=\sum_{i,j}\overline{\dot{E}_{i}(X)
\dot{E}_{j}(X')\delta\biglb(E-E_{i}(X)\bigrb)
\delta\biglb(E'\!-E_{j}(X')\bigrb)}.\label{CXdef}
\end{equation}
Since
\begin{equation}
\dot{A}(X)=\sum_{i=1}^{N}\dot{E}_{i}(X)
\frac{d}{dE_{i}}a\biglb(E_{i}(X)\bigrb),
\end{equation}
one has, upon partial integration,
\begin{equation}
\overline{\dot{A}(X)\dot{A}(X')}=
\int_{-\infty}^{\infty}\! dE\int_{-\infty}^{\infty}\!
dE'\,a(E)a(E')\frac{\partial^{2}}{\partial E\partial E'}
C(E,X,E',X').\label{AdotCXdef}
\end{equation}
Comparison of Eqs.\ (\ref{AdotSXdef}) and (\ref{AdotCXdef}) shows that the
density and current correlation functions $S$ and $C$ are related by
\begin{equation}
\frac{\partial^{2}}{\partial E\partial E'}C(E,X,E',X')=
\frac{\partial^{2}}{\partial X\partial X'}S(E,X,E',X').\label{SCrelation}
\end{equation}

We assume that $S(E,X,E',X+\delta X)\equiv S(E,E',\delta X)$ depends only on
the parameter increment $\delta X$. When considering a particular physical
system, such as the hydrogen atom in a magnetic field, this may require a
rescaling of the energy levels, to eliminate a systematic drift in $E_{i}$
versus $X$ (cf.\ Fig.\ 1).
Since, by definition, $S(E,E',\delta X)=S(E',E,-\delta X)$, the correlators
$\overline{\delta A(X)\delta A(X+\delta X)}$ and
$\overline{\dot{A}(X)\dot{A}(X+\delta X)}$ are even functions of $\delta X$.
Furthermore, we have the sum rule
\begin{equation}
\int_{0}^{\infty}\!d\,\delta X\,\overline{\dot{A}(X)\dot{A}
(X+\delta X)}=0,\label{sumrule}
\end{equation}
for any linear statistic.

In the following sections we will compute the density correlation function
$S(E,E',X)$ from the Brownian-motion model described in Sec.\ \ref{brownian}.
In view of Eq.\ (\ref{SXdef}) and the identification (\ref{tauX}), we have the
relation
\begin{eqnarray}
S(E,E',X)&=&\int_{-\infty}^{\infty}\!dE_{1}^{0}
\cdots\int_{-\infty}^{\infty}\!dE_{N}^{0}\int_{-\infty}^{\infty}
\!dE_{1}\cdots\int_{-\infty}^{\infty}\!dE_{N}\left(\sum_{i,j}
\delta(E-E_{i}^{0})\delta(E'-E_{j})\right)\nonumber\\
&&\mbox{}\times P_{\rm eq}(\{E_{n}^{0}\})
\left(P(\{E_{n}\},X^{2})-P_{\rm eq}(\{E_{n}\})\right)\label{SP}
\end{eqnarray}
between the density correlation function and the solution $P(\{E_{n}\},\tau)$
of the Fokker-Planck equation (\ref{FokkerPlanck}) with initial condition
(\ref{Pinitial}). Once we have $S$, the current correlation function $C$ and
the correlators of $A$ and $\dot{A}$ follow from Eqs.\ (\ref{ASXdef}),
(\ref{AdotSXdef}), and (\ref{SCrelation}).

\section{Asymptotic solution}
\label{asymptotic}

In this section we compute the large-$N$ asymptotic limits of the density and
current correlation functions $S(E,E',X)$ and $C(E,E',X)$. By ``asymptotic'' we
mean that the expressions obtained hold in the limit $N\rightarrow\infty$ in
the energy range $|E-E'|\gg\Delta$ for all $X$ and in the parameter range
$X\gg\Delta\sqrt{\gamma}$ for all $E,E'$. A justification of our asymptotic
analysis will be given in Sec.\ \ref{exact}, when we compare with an exact
result for $\beta=2$.
We assume that the $N\rightarrow\infty$ limit is accompanied by a rescaling of
the confining potential $V(E)$ in Eq.\ (\ref{Wdef}), such that the mean density
of states remains the same. An explicit example of such a rescaling is given in
Sec.\ \ref{exact}.

The first step in the analysis is to reduce the Fokker-Planck equation
(\ref{FokkerPlanck}) to an evolution equation for the average density of
eigenvalues
\begin{equation}
\rho(E,\tau)=\int_{-\infty}^{\infty}\!dE_{1}\cdots
\int_{-\infty}^{\infty}\!dE_{N}\,P(\{E_{n}\},\tau)
\sum_{i=1}^{N}\delta(E-E_{i}).\label{rhodef}
\end{equation}
This problem was solved by Dyson\cite{Dys62} in the limit $N\rightarrow\infty$,
with the result
\begin{equation}
\gamma\frac{\partial}{\partial\tau}\rho(E,\tau)=
\frac{\partial}{\partial E}\left[\rho(E,\tau)
\frac{\partial}{\partial E}\left(\vphantom{\frac{\partial}
{\partial E}}V(E)-\int_{-\infty}^{\infty}\!dE'\,\rho(E',\tau)
\ln|E-E'|\right)\right].\label{DRHODTAU}
\end{equation}
Corrections to Eq.\ (\ref{DRHODTAU}) are smaller by an order $N^{-1}\ln N$. To
the same order, the equilibrium density $\rho_{\rm eq}(E)$ (defined as in Eq.\
(\ref{rhodef}) with $P$ replaced by $P_{\rm eq}$) satisfies\cite{Dys62}
\begin{equation}
\frac{\partial}{\partial E}\left(V(E)-\int_{-\infty}^{\infty}
\!dE'\,\rho_{\rm eq}(E')\ln|E-E'|\right)=0.\label{rhoeq}
\end{equation}
To make this paper selfcontained, we present Dyson's derivation of Eq.\
(\ref{DRHODTAU}) in the Appendix.

The next step is to reduce Eq.\ (\ref{DRHODTAU}) to a diffusion equation by
linearizing $\rho$ around $\rho_{\rm eq}$. This is consistent with the
large-$N$ limit, since $\rho$ is of order $N$ while fluctuations in the density
are of order one.\cite{Dys62} We write $\rho(E,\tau)=\rho_{\rm
eq}(E)+\delta\rho(E,\tau)$ and find, to first order in $\delta\rho$,
\begin{eqnarray}
&&\frac{\partial}{\partial\tau}\delta\rho(E,\tau)=
\frac{\partial}{\partial E}\int_{-\infty}^{\infty}
\!dE'\,D(E,E')\frac{\partial}{\partial E'}\,\delta\rho(E',\tau),
\label{DRHODTAU2}\\
&&D(E,E')=-\gamma^{-1}\rho_{\rm eq}(E)\ln|E-E'|.\label{DE}
\end{eqnarray}
Eq.\ (\ref{DRHODTAU2}) has the form of a non-local diffusion equation, with
diffusion kernel (\ref{DE}).

To proceed we assume a constant density of states $\rho_{\rm
eq}(E)\equiv\rho_{0}\equiv 1/\Delta$ over the energy range of interest (which
is the energy range where the function $a(E)$ in the linear statistic
(\ref{Adef}) differs appreciably from zero). The diffusion kernel can then be
taken to be translationally invariant, $D(E,E')\equiv D(E'-E)$, with Fourier
transform
\begin{equation}
D(k)\equiv\int_{-\infty}^{\infty}\!dE\,{\rm e}^{{\rm i}kE}D(E)=
\frac{\rho_{0}\pi}{\gamma |k|}.\label{Dk}
\end{equation}
Equation (\ref{DRHODTAU2}) becomes an ordinary differential equation in
$k$-space, with solution
\begin{equation}
\delta\rho(k,\tau)=\delta\rho(k,0)\exp[-k^{2}D(k)\tau].\label{deltarhosol}
\end{equation}

In view of Eq.\ (\ref{Pinitial}), the initial condition on the eigenvalue
density is
\begin{equation}
\rho(E,0)=\sum_{i=1}^{N}\delta(E-E_{i}^{0}).\label{rho0}
\end{equation}
We define the equilibrium average $\langle f\rangle_{\rm eq}$ of an arbitrary
function $f(\{E_{n}^{0}\})$ of the initial configuration by
\begin{equation}
\langle f\rangle_{\rm eq}=\int_{-\infty}^{\infty}\!dE_{1}^{0}\cdots
\int_{-\infty}^{\infty}\!dE_{N}^{0}\,P_{\rm eq}(\{E_{n}^{0}\})
f(\{E_{n}^{0}\}).\label{eqavdef}
\end{equation}
Using also definition (\ref{rhodef}), Eq.\ (\ref{SP}) for the density
correlation function $S(E,E',X)$ can be written as
\begin{eqnarray}
S(E,E',X)&=&\langle\rho(E,0)\rho(E',X^{2})\rangle_{\rm eq}-
\rho_{\rm eq}(E)\rho_{\rm eq}(E')\nonumber\\
&=&\langle\delta\rho(E,0)\delta\rho(E',X^{2})\rangle_{\rm eq}.\label{Sdef2}
\end{eqnarray}
In the second equality we have used that $\langle\rho(E,\tau)\rangle_{\rm
eq}=\rho_{\rm eq}(E)$. The correlation function $K(E,E')$ is defined by [cf.\
Eq.\ (\ref{Kdef})]
\begin{equation}
K(E,E')=-\langle\delta\rho(E,0)\delta\rho(E',0)\rangle_{\rm eq}=
-S(E,E',0).\label{Kdef2}
\end{equation}
Over the energy range of a constant density of states, the correlation
functions $S(E,E',X)\equiv S(E'-E,X)$ and $K(E,E')\equiv K(E'-E)$ are
translationally invariant, with Fourier transforms $S(k,X)$ and $K(k)$.
According to Eqs.\ (\ref{deltarhosol}), (\ref{Sdef2}), and (\ref{Kdef2}), we
have
\begin{equation}
S(k,X)=-K(k)\exp[-k^{2}D(k)X^{2}].\label{Sksol}
\end{equation}
The function $K(k)$ is known.\cite{Meh91} In the limit $N\rightarrow\infty$,
one has asymptotically
\begin{equation}
K(k)=-\frac{|k|}{\pi\beta},\label{Kksol}
\end{equation}
{\em independent\/} of $V(E)$.\cite{Bee93} Eq.\ (\ref{Kksol}) is the Fourier
transform of Eq.\ (\ref{K0}), and holds for energy scales $k^{-1}\gg\Delta$
large compared to the mean level spacing. (This is the relevant regime, since
the function $a(E)$ in the linear statistic (\ref{Adef}) is assumed to be
smooth on the scale of the level spacing.)

Combining Eqs.\ (\ref{Dk}), (\ref{Sksol}), and (\ref{Kksol}), we conclude that
the density correlation function is given by
\begin{eqnarray}
&&S(k,X)=\frac{|k|}{\pi\beta}\exp(-\xi^{2}|k|),\label{SkX}\\
&&\xi\equiv X\left(\pi\rho_{0}/\gamma\right)^{1/2}.\label{xidef}
\end{eqnarray}
The $E$-space correlation function becomes, upon inverse Fourier
transformation,
\begin{equation}
S(E,X)=\frac{1}{2\pi^{2}\beta}\,\frac{\partial^{2}}
{\partial E^{2}}\ln(\xi^{4}+E^{2}).\label{SEX}
\end{equation}
The current correlation function $C(E,E',X,X')\equiv C(E'-E,X'-X)$ is obtained
from $S$ by means of relation (\ref{SCrelation}), which in $k$-space takes the
form
\begin{equation}
C(k,X)=-\frac{1}{k^{2}}\,\frac{\partial^{2}}
{\partial X^{2}}S(k,X).\label{SCrelation2}
\end{equation}
We find from Eqs.\ (\ref{SkX}) and (\ref{SCrelation2}):
\begin{eqnarray}
C(k,X)&=&\frac{2\rho_{0}}{\beta\gamma}(1-2\xi^{2}|k|)
\exp(-\xi^{2}|k|),\label{CkX}\\
C(E,X)&=&\frac{1}{2\pi^{2}\beta}\,\frac{\partial^{2}}
{\partial X^{2}}\ln(\xi^{4}+E^{2}).\label{CEX}
\end{eqnarray}

The asymptotic results for the correlation functions given above can be used to
compute the $N\rightarrow\infty$ limit of the integrated correlator
$\chi_{_A}$, defined by Eqs.\ (\ref{chidef}) and (\ref{ASXdef}). The $k$-space
expression for $\chi_{_A}$ is
\begin{eqnarray}
&&\chi_{_A}=\frac{1}{2\pi}\int_{0}^{\infty}
\!dX\int_{-\infty}^{\infty}\!dk\,|a(k)|^{2}S(k,X),\label{chikdef}\\
&&a(k)\equiv\int_{-\infty}^{\infty}
\!dE\,{\rm e}^{{\rm i}kE}a(E).\label{akdef}
\end{eqnarray}
Substituting the asymptotic formula (\ref{SkX}), and carrying out the integral
over $X$, we obtain the result
\begin{equation}
\chi_{_A}=\frac{1}{2\pi^{2}\beta}(\gamma/\rho_{0})^{1/2}
\int_{0}^{\infty}\!dk\,|a(k)|^{2}k^{1/2}.\label{chiksol}
\end{equation}

\section{Exact solution}
\label{exact}

The Fokker-Planck equation (\ref{FokkerPlanck}) can be solved exactly for the
{\em Gaussian ensemble}, which is the case of a parabolic potential
$V(E)=cE^{2}$ ($c$ is an arbitrary positive constant).
The eigenfrequencies and eigenfunctions of the Fokker-Planck equation were
constructed by Sutherland,\cite{Sut72} by mapping it onto a Schr\"{o}dinger
equation. Here we use the same method to compute the correlation functions for
$\beta=2$, and compare with the asymptotic $N\rightarrow\infty$ results of
Sec.\ \ref{asymptotic}.

\subsection{Sutherland's method}
\label{Sutherland}

To map the Fokker-Planck equation (\ref{FokkerPlanck}) onto a Schr\"{o}dinger
equation we substitute
\begin{equation}
P(\{E_{n}\},\tau)={\rm e}^{-\frac{\beta}{2}
W(\{E_{n}\})}\Psi(\{E_{n}\},\tau),\label{subs}
\end{equation}
where $W$ is given by Eq.\ (\ref{Wdef}) with $V(E)=cE^{2}$.
Sutherland\cite{Sut72} used a different mapping (with $\beta$ instead of
$\beta/2$ in the exponent), but this one is more suitable for our purpose.
Substitution of Eq.\ (\ref{subs}) into Eq.\ (\ref{FokkerPlanck}) yields for
$\Psi$ the equation
\begin{equation}
-\frac{\partial\Psi}{\partial\tau}=-\frac{1}{\beta\gamma}
\sum_{i=1}^{N}\frac{\partial^{2}\Psi}{\partial E_{i}^{2}}+
\frac{1}{2\gamma}\Psi\sum_{i=1}^{N}\left[\frac{\beta}{2}
\left(\frac{\partial W}{\partial E_{i}}\right)^{2}-
\frac{\partial^{2}W}{\partial E_{i}^{2}}\right].\label{Schro1}
\end{equation}
The expression between square brackets is evaluated as follows:
\begin{eqnarray}
\sum_{i=1}^{N}\frac{\partial^{2}W}{\partial E_{i}^{2}}&=&
\sum_{i}\sum_{j(\neq i)}\frac{1}{(E_{i}-E_{j})^{2}}+2cN,\\
\sum_{i=1}^{N}\left(\frac{\partial W}{\partial E_{i}}\right)^{2}&=&
\sum_{i}\sum_{j(\neq i)}\sum_{k(\neq i)}\frac{1}{E_{i}-E_{j}}\,
\frac{1}{E_{i}-E_{k}}+4c^{2}\sum_{i}E_{i}^{2}-4c\sum_{i}
\sum_{j(\neq i)}\frac{E_{i}}{E_{i}-E_{j}}\nonumber\\
&=&\sum_{i}\sum_{j(\neq i)}\frac{1}
{(E_{i}-E_{j})^{2}}+4c^{2}\sum_{i}E_{i}^{2}-2cN(N-1).
\end{eqnarray}
In the final equality we have used that for any three distinct indices $i,j,k$
\begin{equation}
\frac{1}{E_{i}-E_{j}}\,\frac{1}{E_{i}-E_{k}}+
\frac{1}{E_{j}-E_{i}}\,\frac{1}{E_{j}-E_{k}}+
\frac{1}{E_{k}-E_{i}}\,\frac{1}{E_{k}-E_{j}}\equiv 0,
\end{equation}
so that the triple sum over $k\neq i\neq j$ collapses to a double sum over
$i\neq j$. Collecting results, we find that $\Psi$ satisfies a Schr\"{o}dinger
equation in imaginary time ($\tau\equiv{\rm i}t$),
\begin{eqnarray}
-\frac{\partial\Psi}{\partial\tau}&=&({\cal H}_{\rm S}-U_{0})
\Psi,\label{Schrodinger}\\
{\cal H}_{\rm S}&=&-\frac{1}{\beta\gamma}\sum_{i}\frac{\partial^{2}}
{\partial E_{i}^{2}}+\frac{\beta-2}{4\gamma}\sum_{i}\sum_{j(\neq i)}
\frac{1}{(E_{i}-E_{j})^{2}}+\frac{\beta c^{2}}{\gamma}\sum_{i}E_{i}^{2},
\label{SutherlandH}\\
U_{0}&=&N\frac{c}{\gamma}+N(N-1)\frac{\beta c}{2\gamma}.\label{U0def}
\end{eqnarray}

The Sutherland Hamiltonian ${\cal H}_{\rm S}$ has an inverse-square interaction
and a parabolic confining potential. The interaction is attractive for
$\beta=1$ and repulsive for $\beta=4$. For $\beta=2$ the interaction vanishes.
Since $\exp(-\beta W)$ is a time-independent solution of the Fokker-Planck
equation (\ref{FokkerPlanck}), $\exp(-\beta W/2)$ is a time-independent
solution of the Schr\"{o}dinger equation (\ref{Schrodinger}) [in view of Eq.\
(\ref{subs})]. Hence once has the eigenvalue equation
\begin{equation}
{\cal H}_{\rm S}\,{\rm e}^{-\frac{\beta}{2}W}=
U_{0}\,{\rm e}^{-\frac{\beta}{2}W}.\label{eigenvalue}
\end{equation}
For a particular ordering of the ``coordinates'' $E_{1},E_{2},\ldots E_{N}$,
the function $\Psi_{0}\propto\exp(-\beta W/2)$ is an eigenfunction of the
$N$-fermion Hamiltonian ${\cal H}_{\rm S}$. Since it is nodeless, it is the
ground state, at energy $U_{0}$. Anti-symmetrization yields the fermion
ground-state wavefunction\cite{Sut72,Sut71}
\begin{equation}
\Psi_{0}(\{E_{n}\})=C{\rm e}^{-\frac{\beta}{2}W(\{E_{n}\})}
\prod_{i<j}\frac{E_{i}-E_{j}}{|E_{i}-E_{j}|},\label{Psi0}
\end{equation}
with $C$ a normalization constant. (Alternatively, we could work with the
symmetric wavefunction $\exp(-\beta W/2)$, which is the ground state for
hard-core bosons.)

We obtain the $N$-particle Green's function $G(\{E_{n}\},\tau)$ of the
Schr\"{o}dinger equation (\ref{Schrodinger}) from $P(\{E_{n}\},\tau)$ by the
similarity transformation
\begin{equation}
G(\{E_{n}\},\tau)={\rm e}^{\frac{\beta}{2}W(\{E_{n}\})}P(\{E_{n}\},\tau)
{\rm e}^{-\frac{\beta}{2}W(\{E_{n}^{0}\})}.\label{similar}
\end{equation}
For $\tau>0$, the function $G$ satisfies
\begin{equation}
-\frac{\partial G}{\partial\tau}=({\cal H}_{\rm S}-U_{0})G,
\label{SchrodingerG}
\end{equation}
in view of Eqs.\ (\ref{subs}) and (\ref{Schrodinger}). The initial condition is
\begin{equation}
G(\{E_{n}\},0)=\prod_{i=1}^{N}\delta(E_{i}^{\vphantom{0}}-E_{i}^{0}),
\label{Ginitial}
\end{equation}
in view of Eq.\ (\ref{Pinitial}). Hence $G$ is indeed a Green's function. In
operator notation,
\begin{equation}
G(\tau)={\rm e}^{-({\cal H}_{\rm S}-U_{0})\tau}.\label{Goperator}
\end{equation}
We note that since the Fokker-Planck equation conserves the ordering of the
levels $E_{1},E_{2},\ldots E_{N}$ for $\tau\geq 0$, we can write Eq.\
(\ref{similar}) equivalently in terms of the anti-symmetrized wavefunction
(\ref{Psi0}),
\begin{equation}
G(\{E_{n}\},\tau)=\Psi_{0}^{-1}(\{E_{n}\})P(\{E_{n}\},\tau)\Psi_{0}
(\{E_{n}^{0}\}).\label{similar2}
\end{equation}

We are now ready to relate the equilibrium density-correlation function in
Dyson's classical Brownian-motion model to the ground-state density-correlation
function in Sutherland's quantum many-body problem in imaginary time. In fact,
we will see that the two correlation functions are {\em identical}. We define
the ground-state expectation value $\langle A\rangle_{0}$ of an operator $A$,
\begin{equation}
\langle A\rangle_{0}=\int_{-\infty}^{\infty}\!
dE_{1}\cdots\int_{-\infty}^{\infty}\!dE_{N}\,\Psi_{0}^{\ast}A
\Psi_{0}^{\vphantom{\ast}}.\label{Oexp}
\end{equation}
The density operator is
\begin{mathletters}
\label{nEtau}
\begin{eqnarray}
&&n(E)=\sum_{i=1}^{N}\delta(E-E_{i}),\label{nEdef}\\
&&n(E,\tau)={\rm e}^{{\cal H}_{\rm S}\tau}n(E)
{\rm e}^{-{\cal H}_{\rm S}\tau},\label{nEtaudef}
\end{eqnarray}
\end{mathletters}%
in the Schr\"{o}dinger and Heisenberg picture, respectively. Combining Eqs.\
(\ref{rhodef}), (\ref{rho0}), (\ref{eqavdef}), and
(\ref{Goperator})--(\ref{nEtau}), one then finds
\begin{eqnarray}
\langle\rho(E',\tau)\rho(E,0)\rangle_{\rm eq}&=&
\langle n(E'){\rm e}^{-({\cal H}_{\rm S}-U_{0})\tau}n(E)\rangle_{0}
\nonumber\\
&=&\langle n(E',\tau)n(E,0)\rangle_{0}.\label{rhonrelation}
\end{eqnarray}
Hence the density correlation function $S(E,E',X)$ defined in Eq.\
(\ref{Sdef2}) is identical to
\begin{equation}
S(E,E',X)=\langle n(E',X^{2})n(E,0)\rangle_{0}-
\langle n(E)\rangle_{0}\langle n(E')\rangle_{0}.\label{SSutherland}
\end{equation}

\subsection{Gaussian unitary ensemble}
\label{Gaussian}

The significance of the formal relationship (\ref{SSutherland}) is that the
quantum mechanical correlator on the right-hand-side can be computed exactly
using the known excitation spectrum of the Sutherland Hamiltonian.\cite{Sut72}
The problem of computing the time-dependent correlation functions of ${\cal
H}_{\rm S}$ was previously considered by Simons, Lee, and Altshuler, in
connection with a microscopic theory of parametric correlations.\cite{Sim93b}
We will return to their work in Sec.\ \ref{microscopic}.
The case $\beta=2$ is particularly simple, since ${\cal H}_{\rm S}$ is then the
Hamiltonian of {\em non-interacting\/} fermions. This is the case of the
Gaussian unitary ensemble.

The single-particle eigenfunctions $\phi_{p}(E)$ and eigenvalues
$\varepsilon_{p}$ of ${\cal H}_{\rm S}$ are (cf.\ Eq.\ (\ref{SutherlandH}) with
$\beta=2$):
\begin{eqnarray}
&&\phi_{p}(E)=(2c/\pi)^{1/4}\,(2^{p}p!)^{-1/2}\,
{\rm e}^{-cE^{2}}{\rm H}_{p}(E\sqrt{2c}),\label{phidef}\\
&&\varepsilon_{p}=(p+\case{1}{2})\omega,\;p=0,1,2,\ldots\label{epsilonm}\\
&&\omega\equiv 2c/\gamma.\label{omegadef}
\end{eqnarray}
The functions ${\rm H}_{p}(x)$ are the Hermite polynomials. The density
operator (\ref{nEtau}) becomes, in second quantization,
\begin{mathletters}
\label{nEtausq}
\begin{eqnarray}
&&n(E)=\sum_{p,q=0}^{\infty}\phi_{p}(E)\phi_{q}(E)c_{p}^{\dagger}
c_{q}^{\vphantom{\dagger}},\label{nEdefsq}\\
&&n(E,\tau)=\sum_{p,q=0}^{\infty}\phi_{p}(E)\phi_{q}(E)
{\rm e}^{(\varepsilon_{p}-\varepsilon_{q})\tau}c_{p}^{\dagger}
c_{q}^{\vphantom{\dagger}},\label{nEtaudefsq}
\end{eqnarray}
\end{mathletters}%
where $c_{p}^{\dagger}$ and $c_{p}$ are fermion creation and annihilation
operators in state $p$. The average density in the $N$-fermion ground state is
\begin{equation}
\langle n(E)\rangle_{0}\equiv\rho_{\rm eq}(E)=
\sum_{p=0}^{N-1}\phi_{p}^{2}(E).\label{naverage}
\end{equation}
To compute the density fluctuations, we need the ground-state expectation value
\begin{equation}
\langle n(E',\tau)n(E,0)\rangle_{0}=\sum_{p',q',p,q=0}^{\infty}
\phi_{p'}(E')\phi_{q'}(E')\phi_{p}(E)\phi_{q}(E)
{\rm e}^{(\varepsilon_{p'}-\varepsilon_{q'})\tau}\langle
c_{p'}^{\dagger}c_{q'}^{\vphantom{\dagger}}c_{p}^{\dagger}
c_{q}^{\vphantom{\dagger}}\rangle_{0}.\label{nncccc}
\end{equation}
The average of the product of four $c$'s evaluates to
\begin{equation}
\langle c_{p'}^{\dagger}c_{q'}^{\vphantom{\dagger}}c_{p}^{\dagger}
c_{q}^{\vphantom{\dagger}}\rangle_{0}=\delta_{pq}\delta_{p'q'}
\theta(N-1-p')\theta(N-1-p)+\delta_{pq'}\delta_{p'q}\theta(N-1-q)
\theta(p-N),\label{cccc}
\end{equation}
where $\delta_{pq}$ is the Kronecker delta and the function $\theta(x)$ equals
1 if $x\geq 0$ and 0 if $x<0$. Collecting results, we find for the density
correlation function (\ref{SSutherland}) the formula
\begin{equation}
S(E,E',X)=\sum_{p=N}^{\infty}\sum_{q=0}^{N-1}
\phi_{p}(E)\phi_{p}(E')\phi_{q}(E)\phi_{q}(E'){\rm e}^{(\varepsilon_{q}-
\varepsilon_{p})X^{2}}.\label{SSutherland2}
\end{equation}

The infinite series over $p$ in Eq.\ (\ref{SSutherland2}) can be reduced to a
finite sum by using an addition theorem for Hermite polynomials:
\begin{eqnarray}
G_{0}(E,E',\tau)&\equiv&\sum_{p=0}^{\infty}\phi_{p}(E)\phi_{p}(E')
{\rm e}^{-\varepsilon_{p}\tau}\nonumber\\
&=&\left(\frac{c}{\pi\sinh\omega\tau}\right)^{1/2}\exp\left(\frac{c}
{\sinh\omega\tau}\left[2EE'-(E^{2}+E'^{\,2})\cosh\omega\tau\right]\right).
\label{GreenHO}
\end{eqnarray}
This is the familiar result for the (imaginary time) Green's function of a
one-dimensional harmonic oscillator (with coordinate $E$, mass $\gamma$, and
oscillator frequency $\omega$). Substitution into Eq.\ (\ref{SSutherland2})
yields
\begin{eqnarray}
S(E,E',X)&=&G_{0}(E,E',X^{2})\sum_{q=0}^{N-1}
\phi_{q}(E)\phi_{q}(E'){\rm e}^{\varepsilon_{q}X^{2}}\nonumber\\
&&\mbox{}-\sum_{p=0}^{N-1}\sum_{q=0}^{N-1}
\phi_{p}(E)\phi_{p}(E')\phi_{q}(E)\phi_{q}(E'){\rm e}^{(\varepsilon_{q}-
\varepsilon_{p})X^{2}}.\label{SSutherland3}
\end{eqnarray}
For $X=0$ the function $G_{0}$ becomes a delta function, so that Eq.\
(\ref{SSutherland3}) reduces to
\begin{equation}
S(E,E',0)=\delta(E-E')\sum_{p=0}^{N-1}\phi_{p}^{2}(E)-\left(
\sum_{p=0}^{N-1}\phi_{p}(E)\phi_{p}(E')\right)^{2}.\label{SSutherland4}
\end{equation}
Eq.\ (\ref{SSutherland4}) was obtained by Mehta\cite{Meh91} using an approach
known as the ``method of orthogonal polymials''. Our Eq.\ (\ref{SSutherland3})
extends this exact result to parametric correlations.

\subsection{Large-$N$ limit}
\label{largeN}

It is instructive to see how the result (\ref{SkX}) of the asymptotic analysis
in Sec.\ \ref{asymptotic} follows (for $\beta=2$) from the large-$N$ limit of
the exact result (\ref{SSutherland2}) for the Gaussian unitary ensemble.

We wish to evaluate the density correlation function (\ref{SSutherland2}) in
the limit $N\rightarrow\infty$, $c\rightarrow 0$, while the product $cN$
remains constant (to ensure a constant density of states, see below). Using the
asymptotic form of the Hermite polynomials ${\rm H}_{p}$ for $p\gg 1$, one has
for the eigenfunctions (\ref{phidef}) the large-$p$ expressions
\begin{mathletters}
\label{phiasymp}
\begin{eqnarray}
\phi_{2p}(E)&=&(-1)^{p}\,(2c/p\pi^{2})^{1/4}\,{\rm e}^{-cE^{2}}
\cos(E\sqrt{8cp}),\label{phieven}\\
\phi_{2p+1}(E)&=&(-1)^{p}\,(2c/p\pi^{2})^{1/4}\,{\rm e}^{-cE^{2}}
\sin(E\sqrt{8cp}).\label{phiodd}
\end{eqnarray}
\end{mathletters}
We need to compute the series
\begin{eqnarray}
\sum_{p=0}^{N-1}\phi_{p}(E)\phi_{p}(E'){\rm e}^{\varepsilon_{p}X^{2}}&=&
\sum_{p=0}^{\frac{1}{2}(N-1)}\phi_{2p}(E)\phi_{2p}(E'){\rm e}^{(2p+
\frac{1}{2})\omega X^{2}}\nonumber\\
&&\mbox{}+\sum_{p=0}^{\frac{1}{2}(N-2)}\phi_{2p+1}(E)\phi_{2p+1}(E')
{\rm e}^{(2p+\frac{3}{2})\omega X^{2}},\label{series1}
\end{eqnarray}
in the limit $N\rightarrow\infty$, $c\rightarrow 0$ at constant $cN$. Note that
$c\rightarrow 0$ implies $\omega\rightarrow 0$, in view of Eq.\
(\ref{omegadef}). Combining Eqs.\ (\ref{phiasymp}) and (\ref{series1}), and
replacing the sum over $p$ by an integral, we find in this limit
\begin{equation}
\sum_{p=0}^{N-1}\phi_{p}(E)\phi_{p}(E'){\rm e}^{\varepsilon_{p} X^{2}}=
\rho_{0}\int_{0}^{1}\!ds\,{\rm e}^{\alpha X^{2} s^{2}}
\cos\biglb(\pi\rho_{0}(E-E')s\bigrb),\label{series2}
\end{equation}
with the definitions
\begin{eqnarray}
&&\rho_{0}\equiv\frac{2}{\pi}(cN)^{1/2},\label{rho0def}\\
&&\alpha\equiv N\omega\equiv\frac{\pi^{2}\rho_{0}^{2}}{2\gamma}.
\label{alphadef}
\end{eqnarray}
Similarly,
\begin{equation}
\sum_{p=N}^{\infty}\phi_{p}(E)\phi_{p}(E'){\rm e}^{-\varepsilon_{p} X^{2}}=
\rho_{0}\int_{1}^{\infty}\!ds\,{\rm e}^{-\alpha X^{2} s^{2}}
\cos\biglb(\pi\rho_{0}(E-E')s\bigrb).\label{series3}
\end{equation}

Substitution of Eq.\ (\ref{series2}) into Eq.\ (\ref{naverage}) gives
\begin{equation}
\rho_{\rm eq}(E)=\rho_{0},\label{rho0asymp}
\end{equation}
justifying the identification (\ref{rho0def}). The limit $N\rightarrow\infty$
yields a uniform density of states in any {\em fixed\/} energy range. At finite
$N$, the density $\rho_{\rm eq}(E)$ vanishes for $\rho_{0}|E|\gtrsim 2N/\pi$,
as follows from a more accurate evaluation of Eq.\
(\ref{naverage}).\cite{Meh91}

Substitution of Eqs.\ (\ref{series2}) and (\ref{series3}) into Eqs.\
(\ref{SSutherland2}) gives an integral expression for the density correlation
function $S(E,E',X)\equiv S(E'-E,X)$,
\begin{equation}
S(E,X)=\rho_{0}^{2}\int_{0}^{1}\!ds\int_{1}^{\infty}\!ds'\,
\exp[\alpha X^{2}(s^{2}-s'^{\,2})]\cos(\pi\rho_{0}Es)
\cos(\pi\rho_{0}Es').\label{Sasymp1}
\end{equation}
The Fourier transform $S(k,X)\equiv\int dE\,S(E,X)\exp({\rm i}kE)$ with respect
to the energy increment becomes
\begin{eqnarray}
&&S(k,X)=\frac{\rho_{0}}{\xi^{2}|k|}\exp(-\xi^{2}|k|q_{\rm max})
\sinh(\xi^{2}|k|q_{\rm min}),\label{Sasymp2}\\
&&q_{\rm min}\equiv{\rm min}\left(1,\frac{|k|}{2\pi\rho_{0}}\right),\;
q_{\rm max}\equiv{\rm max}\left(1,\frac{|k|}{2\pi\rho_{0}}\right).
\label{qminmax}
\end{eqnarray}
The variable $\xi$ was defined in Eq.\ (\ref{xidef}).

The result (\ref{Sasymp2}) holds in the large-$N$ limit (at constant density of
states) in any fixed $k$-range. We now further restrict ourselves to energy
scales bigger than the mean level spacing $\Delta\equiv\rho_{0}^{-1}$, i.e.\ to
the range $k\ll\rho_{0}$. Eq.\ (\ref{Sasymp2}) then simplifies to
\begin{equation}
S(k,X)=\frac{|k|}{2\pi}\exp(-\xi^{2}|k|),\label{Sasymp3}
\end{equation}
in agreement with Eq.\ (\ref{SkX}) for $\beta=2$. The correlation function
$S(k,X)$ is only appreciably different from zero if $\xi^{2}|k|\lesssim 1$.
Hence the restriction $k\ll\rho_{0}$ on Eq.\ (\ref{Sasymp3}) becomes irrelevant
if $\xi^{2}\rho_{0}\gg 1$. This implies that the asymptotic expressions for the
correlation functions $S(E,X)$ [and $C(E,X)$] of Sec.\ \ref{asymptotic} hold
for $N\rightarrow\infty$ in the energy range $E\gg\Delta$ for all $X$ and in
the parameter range $X\gg\Delta\sqrt{\gamma}$ for all $E$.
If {\em both\/} $E\lesssim\Delta$ {\em and\/} $X\lesssim\Delta\sqrt{\gamma}$
one can not use Eq.\ (\ref{Sasymp3}), but should use instead the full
expressions (\ref{Sasymp1}) or (\ref{Sasymp2}).

Once we have the density correlation function $S(k,X)$, the current correlation
function $C(k,X)$ follows directly in view of the relation (\ref{SCrelation2}).
{}From Eq.\ (\ref{Sasymp2}) one thus finds
\begin{eqnarray}
C(k,X)&=&\frac{2\pi\rho_{0}^{2}}{\gamma\xi^{4}|k|^{3}}
\exp(-\xi^{2}|k|q_{\rm max})\left(\xi^{2}|k|q_{\rm min}
\cosh(\xi^{2}|k|q_{\rm min})[3+4\xi^{2}|k|q_{\rm max}]\right.\nonumber\\
&&\left.\mbox{}-\sinh(\xi^{2}|k|q_{\rm min})
[3+3\xi^{2}|k|q_{\rm max}+2\xi^{4}k^{2}(q_{\rm max}^{2}+q_{\rm min}^{2})]
\right).\label{Casymp2}
\end{eqnarray}
Eq.\ (\ref{Casymp2}) holds for $N\rightarrow\infty$ and any $k$. For
$k\ll\rho_{0}$ it reduces to the asymptotic expression (\ref{CkX}) of Sec.\
\ref{asymptotic} (with $\beta=2$).

\section{Extension to multiple parameters}
\label{multiple}

In this section we show how the Brownian-motion model of Sec.\ \ref{brownian}
can be extended to a parameter {\em vector\/} $\vec{X}=X_{1},X_{2},\ldots
X_{d}$, relevant for a statistical description of the dispersion relation of a
$d$-dimensional crystalline lattice.\cite{Sim93} The Brownian motion of the
energy levels $E_{n}(\vec{X})$ then takes place in a fictitious world with
multiple temporal dimensions $\vec{\tau}=\tau_{1},\tau_{2},\ldots \tau_{d}$.

We assume that any systematic drift in the energy levels is eliminated by a
rescaling, so that
\begin{equation}
\overline{\partial_{\mu}E_{n}(\vec{X})}=0.
\end{equation}
(We abbreviate $\partial_{\mu}\equiv\partial/\partial X_{\mu}$.) We also assume
that the different parameters $X_{\mu}$ are independent, that is to say
\begin{equation}
\overline{\partial_{\mu}E_{n}(\vec{X})\partial_{\nu}E_{n}(\vec{X})}=0
\;{\rm if}\; \mu\neq\nu.
\end{equation}
Let $\vec{\tau}=0$ coincide with $\vec{X}=0$. The initial condition on the
distribution function $P(\{E_{n}\},\vec{\tau})$ is
\begin{equation}
P(\{E_{n}\},\vec{\tau}=0)=\prod_{i=1}^{N}\delta(E_{i}^{\vphantom{0}}-
E_{i}^{0}),\label{Pinitial2}
\end{equation}
with $E_{i}^{0}$ the eigenvalues of ${\cal H}(\vec{X}=0)$. For $\tau_{\mu}>0$
the distribution function evolves according to the multiple-time-dimensional
generalization of the Fokker-Planck equation (\ref{FokkerPlanck}),
\begin{equation}
\frac{1}{d}\sum_{\mu=1}^{d}\gamma_{\mu}\frac{\partial P}
{\partial\tau_{\mu}}=\sum_{i=1}^{N}\frac{\partial}
{\partial E_{i}}\left(P\frac{\partial W}{\partial E_{i}}+\beta^{-1}
\frac{\partial P}{\partial E_{i}}\right).\label{FokkerPlanck2}
\end{equation}
By comparing the initial average rate of change of the energy levels as a
function of $\vec{X}$ and $\vec{\tau}$,
\begin{eqnarray}
\overline{(E_{i}(\vec{X})-E_{i}^{0})^{2}}&=&\sum_{\mu=1}^{d}X_{\mu}^{2}\,
\overline{(\partial_{\mu}E_{i})^{2}}+{\cal O}(X^{3}),\\
\overline{(E_{i}(\vec{\tau})-E_{i}^{0})^{2}}&=&\frac{2}{\beta}
\sum_{\mu=1}^{d}\frac{\tau_{\mu}}{\gamma_{\mu}}+{\cal O}(\tau^{2}),
\end{eqnarray}
we arrive as in Sec.\ \ref{brownian} at the identifications
\begin{eqnarray}
\tau_{\mu}&=&X_{\mu}^{2},\label{tauX2}\\
2/\beta\gamma_{\mu}&=&\overline{(\partial_{\mu}E_{i})^{2}}.
\label{gammarelation2}
\end{eqnarray}

The Fokker-Planck equation (\ref{FokkerPlanck2}) can now be reduced to a
non-local diffusion equation as in Sec.\ \ref{asymptotic},
\begin{equation}
\frac{1}{d}\sum_{\mu=1}^{d}\gamma_{\mu}\frac{\partial}
{\partial\tau_{\mu}}\delta\rho(E,\vec{\tau})=-\frac{\partial}
{\partial E}\int_{-\infty}^{\infty}\!dE'\,\rho_{\rm eq}(E)
\ln|E-E'|\frac{\partial}{\partial E'}\,\delta\rho(E',\vec{\tau}),
\label{rhotaumu}
\end{equation}
valid asymptotically for $N\rightarrow\infty$. For a constant density of states
$\rho_{0}$ the diffusion kernel becomes translationally invariant. Equation
(\ref{rhotaumu}) then has the $k$-space solution
\begin{equation}
\delta\rho(k,\vec{\tau})=\delta\rho(k,0)\exp\left[
-\pi\rho_{0}|k|\sum_{\mu=1}^{d}\tau_{\mu}/\gamma_{\mu}
\right],\label{deltarhosol2}
\end{equation}
which implies for the density correlation function [cf.\ Eq.\ (\ref{SkX})]
\begin{equation}
S(k,\vec{X})=\frac{|k|}{\pi\beta}\exp\left[-\pi\rho_{0}|k|
\sum_{\mu=1}^{d}X_{\mu}^{2}/\gamma_{\mu}\right].\label{SkXmu}
\end{equation}
Here we have also used the identification (\ref{tauX2}). The $E$-space
correlation function becomes, upon inverse Fourier transformation,
\begin{equation}
S(E,\vec{X})=\frac{1}{2\pi^{2}\beta}\,\frac{\partial^{2}}
{\partial E^{2}}\ln\left(E^{2}+[\pi\rho_{0}|k|
\sum_{\mu=1}^{d}X_{\mu}^{2}/\gamma_{\mu}]^{2}\right).\label{SEXmu}
\end{equation}

The current correlation function
\begin{equation}
C_{\mu\nu}(E,\vec{X},E',\vec{X}')=\sum_{i,j}\overline{\biglb(
\partial_{\mu}E_{i}(\vec{X})\bigrb)\biglb(\partial_{\nu}E_{j}
(\vec{X}')\bigrb)\delta\biglb(E-E_{i}(\vec{X})\bigrb)
\delta\biglb(E'\!-E_{j}(\vec{X}')\bigrb)}\label{CmunuXdef}
\end{equation}
is related to the density correlation function $S(E,\vec{X},E',\vec{X}')$ by
\begin{equation}
\frac{\partial^{2}}{\partial E\partial E'}
C_{\mu\nu}(E,\vec{X},E',\vec{X}')=
\frac{\partial^{2}}{\partial X_{\mu}\partial X'_{\nu}}S(E,\vec{X},E',
\vec{X}').\label{SCmunurelation}
\end{equation}
Because of translational invariance, $C_{\mu\nu}(E,\vec{X},E',\vec{X}')=
C_{\mu\nu}(E'-E,\vec{X}'-\vec{X})$, $S(E,\vec{X},E',\vec{X}')=
S(E'-E,\vec{X}'-\vec{X})$. In $k$-space Eq.\ (\ref{SCmunurelation}) then takes
the form
\begin{equation}
C_{\mu\nu}(k,\vec{X})=-\frac{1}{k^{2}}\,\frac{\partial^{2}}
{\partial X_{\mu}X_{\nu}}S(k,\vec{X}).\label{SCmunurelation2}
\end{equation}
{}From Eqs.\ (\ref{SkXmu}) and (\ref{SCmunurelation2}) we find for the current
correlation function the $k$ and $E$-space expressions
\begin{eqnarray}
C_{\mu\nu}(k,\vec{X})&=&\frac{2\rho_{0}}{\beta}\left(
\frac{\delta_{\mu\nu}}{\gamma_{\mu}}-2\pi\rho_{0}|k|
\frac{X_{\mu}X_{\nu}}{\gamma_{\mu}\gamma_{\nu}}\right)
\exp\left[-\pi\rho_{0}|k|\sum_{\lambda=1}^{d}X_{\lambda}^{2}/
\gamma_{\lambda}\right],\label{CkXmunu}\\
C_{\mu\nu}(E,\vec{X})&=&\frac{1}{2\pi^{2}\beta}\,\frac{\partial^{2}}
{\partial X_{\mu}X_{\nu}}\ln\left(E^{2}+[\pi\rho_{0}|k|\sum_{\mu=1}^{d}
X_{\mu}^{2}/\gamma_{\mu}]^{2}\right).\label{CEXmunu}
\end{eqnarray}

For $d=1$ the correlation functions (\ref{SEXmu}) and (\ref{CEXmunu}) reduce to
the results (\ref{SEX}) and (\ref{CEX}) of Sec.\ \ref{asymptotic}.

\section{Comparison with microscopic theory}
\label{microscopic}

\subsection{Diagrammatic perturbation theory}
\label{diagrammatic}

The asymptotic analysis of Sec.\ \ref{asymptotic} yields the density and
current correlation functions in the limit $N\rightarrow\infty$ if
$E\gg\Delta\equiv 1/\rho_{0}$ for all $X$ and if $X\gg X_{\rm
c}\equiv\Delta\sqrt{\gamma}$ for all $E$. We will now show that the
random-matrix theory (RMT) in this regime agrees with the diagrammatic
perturbation theory of Szafer, Altshuler, and Simons.\cite{Sza93,Sim93}

When $E\rightarrow 0$, our result (\ref{CEX}) for the current correlation
function $C(E,X)$ reduces to
\begin{eqnarray}
C(0,X)&=&\frac{2}{\pi^{2}\beta}\,\frac{\partial^{2}}
{\partial X^{2}}\ln|X|\nonumber\\
&=&-\frac{2}{\pi^{2}\beta X^{2}}, \;{\rm if}\;X\neq 0,\label{C0X}
\end{eqnarray}
{\em independent\/} of the microscopic parameters $\rho_{0}$ and $\gamma$. Eq.\
(\ref{C0X}), obtained here from RMT, is precisely the universal correlator
(\ref{C0}) which Szafer and Altshuler\cite{Sza93} derived from diagrammatic
perturbation theory.

At $X=0$, the function $C(0,X)$ according to Eq.\ (\ref{C0X}) has an integrable
singularity consisting of a positive peak such that the integral over all $X$
vanishes. This is a special case of the general sumrule
\begin{equation}
\int_{0}^{\infty}\!dX\,C(E,X)=0,\label{sumrule2}
\end{equation}
which follows from Eq.\ (\ref{CEX}) [cf.\ also Eq.\ (\ref{sumrule})]. The peak
of positive correlation has infinitesimal width in the limit $E\rightarrow 0$.
At non-zero $E$ the peak has a finite width of order $X_{\rm
c}(E/\Delta)^{1/2}$, as illustrated in Fig.\ 2, where we have plotted $C(E,X)$
from Eq.\ (\ref{CEX}) for $E=0.1\,\Delta$ (dashed curve).

As discussed in Sec.\ \ref{exact}, the asymptotic formula (\ref{C0X}) becomes
exact only for $X\gg X_{\rm c}$. [Compare with the solid curve in Fig.\ 2,
computed from the exact result (\ref{Casymp2}).] Using the definition of the
generalized Thouless energy\cite{Sim93}
\begin{equation}
{\cal E}_{\rm c}\equiv\Delta^{-1}\overline{\dot{E}_{i}^{2}},\label{Ecdef}
\end{equation}
and the relationship (\ref{gammarelation}) between $\gamma$ and
$\overline{\dot{E}_{i}^{2}}$, one can write
\begin{equation}
X_{\rm c}\equiv\Delta\sqrt{\gamma}\equiv\left(
\frac{2\Delta}{\beta{\cal E}_{\rm c}}\right)^{1/2}.\label{Xcdef}
\end{equation}
In Ref.\ \onlinecite{Sza93} the parameter $X$ is the magnetic flux increment in
units of $h/e$. Then ${\cal E}_{\rm c}$ is the conventional Thouless
energy\cite{Akk92} $E_{\rm c}\simeq\hbar v_{\rm F}l/L^{2}$, related to the
conductance $g$ (in units of $e^{2}/h$) by $g\simeq E_{\rm c}/\Delta$.
The Aharonov-Bohm periodicity implies in this case the additional restriction
$X\ll 1$ to Eq.\ (\ref{C0X}) (which is compatible with the condition $X\gg
X_{\rm c}$ because $X_{\rm c}\simeq g^{-1/2}\ll 1$ in the metallic regime).

We have shown that the $E\rightarrow 0$ limit of $C(E,X)$ obtained from RMT
agrees with the microscopic theory. What about non-zero energy differences?
This is most easily discussed in terms of the density correlation function
$S(E,X)$, to which $C(E,X)$ is related via Eq.\ (\ref{SCrelation2}). Using
$\xi\equiv X(\frac{1}{2}\pi\beta{\cal E}_{\rm c})^{1/2}$, we find that the
result (\ref{SEX}) can be rewritten identically as
\begin{equation}
S(E,X)=\frac{1}{\pi^{2}\beta}{\rm Re}\left({\rm i}E+
\case{1}{2}\pi\beta{\cal E}_{\rm c}X^{2}\right)^{-2},\label{SEX2}
\end{equation}
which agrees with the diagrammatic perturbation theory\cite{Sza93,Sim93}
provided $E\ll E_{\rm c}$. The deviation between RMT and the microscopic theory
on energy scales greater than the Thouless energy $E_{\rm c}$ is well known
from the work by Altshuler and Shklovski\u{\i} on parameter-independent
correlations.\cite{Alt86}

One can similarly show that the correlation functions for {\em multiple\/}
parameters $X_{\mu}$ ($\mu=1,2,\ldots d$), obtained from RMT in Sec.\
\ref{multiple}, agree with the results which Simons and Altshuler\cite{Sim93}
obtained by microscopic theory. In particular, we find from Eq.\
(\ref{CEXmunu}) that
\begin{equation}
\sum_{\mu=1}^{d}\gamma_{\mu}C_{\mu\mu}(0,\vec{X})=
\frac{2d-4}{\pi^{2}\beta}[\sum_{\mu=1}^{d}X_{\mu}^{2}/
\gamma_{\mu}]^{-1},\label{C0Xmu}
\end{equation}
in agreement with Ref.\ \onlinecite{Sim93}. The correlator (\ref{C0Xmu}) is the
multiple-parameter generalization of the universal correlator (\ref{C0}).
Simons and Altshuler have discussed the physical origin of the different sign
of the correlator for $d<2$ and $d>2$.

\subsection{Non-linear sigma model}
\label{sigma}

The restriction on the asymptotic analysis that either $X\gg X_{\rm c}$ or
$E\gg\Delta$ is removed by the exact solution of Sec.\ \ref{exact} for the
Gaussian unitary ensemble ($\beta=2$). The density correlation function in the
limit $N\rightarrow\infty$ is given for this random-matrix ensemble by Eq.\
(\ref{Sasymp1}) in $E$-space and by Eq.\ (\ref{Sasymp2}) in $k$-space. In terms
of the generalized Thouless energy (\ref{Ecdef}), the $E$-space expression can
be written as
\begin{equation}
S(E,X)=\Delta^{-2}\int_{0}^{1}\!ds\int_{1}^{\infty}\!ds'\,
\exp\left[\frac{\pi^{2}{\cal E}_{\rm c}}{2\Delta}X^{2}
(s^{2}-s'^{\,2})\right]\cos(\pi sE/\Delta)\cos(\pi s'E/\Delta).
\label{Sasymp1a}
\end{equation}
This is precisely the result of the microscopic theory of Simons and
Altshuler.\cite{Sim93} If either $X\gg X_{\rm c}$ or $E\gg\Delta$, Eq.\
(\ref{Sasymp1a}) reduces to Eq.\ (\ref{SEX}) with $\beta=2$. Simons and
Altshuler were able to extend the microscopic theory to the regime $X\lesssim
X_{\rm c}$, $E\lesssim\Delta$, which is not obtainable by perturbation theory,
by using a supersymmetry formulation followed by a mapping onto a non-linear
sigma model.\cite{Efe83}
As emphasized by these authors, it is quite remarkable that the microscopic
parameters enter only via the quantities $\Delta$ and ${\cal E}_{\rm c}$, so
that a rescaling of the $E$ and $X$ variables maps all density correlation
functions onto a single universal function.

Simons and Altshuler have also computed the small $E$ and $X$ behavior of
$S(E,X)$ from the microscopic theory in the presence of time-reversal symmetry,
i.e.\ for $\beta=1$. Again, they used a mapping onto a non-linear sigma model
to go beyond perturbation theory.
We have no RMT result for the small $E$ and $X$ behavior in this case, which
would correspond to the orthogonal ensemble. The case $\beta=4$ of strong
spin-orbit scattering (symplectic ensemble in RMT) has not yet been treated by
microscopic theory, and only in the asymptotic limit by RMT.

In Ref.\ \onlinecite{Sim93b}, Simons, Lee, and Altshuler have argued
convincingly (although not completely proven) that the density correlation
functions of the non-linear sigma model and the Sutherland Hamiltonian are
equivalent for $\beta=1,2$, and 4. In the present paper, in Sec.\ \ref{exact},
we have proven the equivalence of the density correlation functions of Dyson's
Brownian-motion model and the Sutherland Hamiltonian.
Taken together, this is evidence for the complete equivalence of the non-linear
sigma model and RMT, although the cases $\beta=1$ and $\beta=4$ still lack a
complete proof.

\subsection{Conclusion}
\label{conclusion}

We have studied the response to an external perturbation of the energy levels
of a quantum mechanical system, by means of the Brownian-motion model
introduced by Dyson in the theory of random matrices. Our results for the
energy and parameter-dependent level-density and current-density correlation
functions $S(E,X)$ and $C(E,X)$ agree with the microscopic theory for a
disordered metallic particle, for energy scales below the Thouless energy
$E_{\rm c}$. This establishes the validity of Dyson's basic assumption,
that parametric correlations are dominated by level repulsion and therefore
solely dependent on the symmetry of the Hamiltonian.

It is likely that the approach developed in this paper can also be used to
describe parametric correlations in random transmission matrices. The analogue
of level repulsion for the transmission eigenvalues is known,\cite{Mut87} and
leads to a pair correlation function $K(T,T')$ which differs from Eq.\
(\ref{K0}) for $K(E,E')$ but has the same universal
$\beta$-dependence.\cite{Bee93}
This suggests that the analogue of the universal correlator (\ref{C0}) exists
as well for the transmission eigenvalues, with obvious implications for the
conductance of a mesoscopic system.

\acknowledgments
Valuable discussions and correspondence with B. L. Altshuler and B. D. Simons
are gratefully acknowledged. This research was supported financially by the
``Ne\-der\-land\-se or\-ga\-ni\-sa\-tie voor We\-ten\-schap\-pe\-lijk
On\-der\-zoek'' (NWO) and by the ``Stich\-ting voor Fun\-da\-men\-teel
On\-der\-zoek der Ma\-te\-rie'' (FOM).

\appendix
\section*{Derivation of Equation (\protect\ref{DRHODTAU})}

For completeness, we present here Dyson's derivation\cite{Dys62} of the
non-linear diffusion equation (\ref{DRHODTAU}) from the Fokker-Planck equation
(\ref{FokkerPlanck}), in the limit $N\rightarrow\infty$.

We multiply Eq.\ (\ref{FokkerPlanck}) by $\delta(E-E_{i})$, integrate over
$E_{1},E_{2},\ldots E_{N}$, and sum over $i$. The result is
\begin{equation}
\gamma\frac{\partial}{\partial\tau}\rho(E,\tau)=
\frac{\partial}{\partial E}\left[\beta^{-1}
\frac{\partial }{\partial E}\rho(E,\tau)+\int_{-\infty}^{\infty}\!
dE_{1}\cdots\int_{-\infty}^{\infty}\! dE_{N}\,P(\{E_{n}\},\tau)
\sum_{i=1}^{N}\delta(E-E_{i})\frac{\partial W}{\partial E_{i}}\right],
\label{int1}
\end{equation}
where $\rho(E,\tau)$ is defined in Eq.\ (\ref{rhodef}). Substitution of the
definition (\ref{Wdef}) of $W(\{E_{n}\})$ into Eq.\ (\ref{int1}) leads to
\begin{equation}
\gamma\frac{\partial}{\partial\tau}\rho(E,\tau)=
\frac{\partial}{\partial E}\left[\beta^{-1}\frac{\partial }
{\partial E}\rho(E,\tau)+\rho(E,\tau)\frac{d}{dE}V(E)-
{\cal P}\!\int_{-\infty}^{\infty}\! dE'\,\frac{\rho_{2}(E,E',\tau)}
{E-E'}\right],\label{int2}
\end{equation}
where ${\cal P}\!\int$ indicates the principal value of the integral. The pair
density $\rho_{2}(E,E',\tau)$ is defined by
\begin{equation}
\rho_{2}(E,E',\tau)=\int_{-\infty}^{\infty}\! dE_{1}
\cdots\int_{-\infty}^{\infty}\! dE_{N}\,P(\{E_{n}\},\tau)
\sum_{i\neq j}\delta(E-E_{i})\delta(E'-E_{j}).
\label{rho2xmean}
\end{equation}
The pair density is symmetric in the energy arguments,
$\rho_{2}(E,E',\tau)=\rho_{2}(E',E,\tau)$, and satisfies the normalization
\begin{equation}
\int_{-\infty}^{\infty}\! dE'\rho_{2}(E,E',\tau)
=(N-1)\rho(E,\tau).\label{normal}
\end{equation}

Following Ref.\ \onlinecite{Dys62} we decompose the pair density into a
correlated and an uncorrelated part,
\begin{equation}
\rho_{2}(E,E',\tau)=
\rho(E,\tau)\rho(E',\tau)
[1-y(E,E',\tau)].\label{ydef}
\end{equation}
The function $y(E,E',\tau)=y(E',E,\tau)$ is symmetric in $E$ and $E'$, and
satisfies
\begin{equation}
\int_{-\infty}^{\infty}\! dE'\,y(E,E',\tau)
\rho(E',\tau)=1,\label{normal1}
\end{equation}
in view of the normalization (\ref{normal}). Substitution of the definition
(\ref{ydef}) into Eq.\ (\ref{int2}) leads to
\begin{eqnarray}
\gamma\frac{\partial}{\partial\tau}\rho(E,\tau)=
\frac{\partial}{\partial E}\left[\beta^{-1}
\frac{\partial }{\partial E}\rho(E,\tau)+
\rho(E,\tau)\frac{\partial }{\partial E}[V(E)+U(E,\tau)]+
\rho(E,\tau)I(E,\tau)\right],\label{int3}
\end{eqnarray}
with the definitions
\begin{eqnarray}
I(E,\tau)&=&{\cal P}\!\int_{-\infty}^{\infty}\!
dE'\rho(E',\tau)\frac{y(E,E',\tau)}{E-E'},\label{Idef1}\\
U(E,\tau)&=&-\int_{-\infty}^{\infty}\!
dE'\rho(E',\tau)\ln|E-E'|.
\label{Udef}
\end{eqnarray}

Eq.\ (\ref{int3}) is still exact. To introduce the approximation we need one
further piece of notation. We re-express the function $y(E,E',\tau)$ in terms
of the sum and difference variables $t=\case{1}{2}(E+E')$ and $s=E'-E$:
\begin{equation}
y(E,E',\tau)=Y\biglb(\case{1}{2}(E+E'),E'-E,\tau\bigrb)
\equiv Y(t,s,\tau).\label{Ydef}
\end{equation}
The function $Y(t,s,\tau)=Y(t,-s,\tau)$ is even in $s$. The normalization
(\ref{normal1}) becomes
\begin{equation}
\int_{-\infty}^{\infty}\! ds\,Y(E+\case{1}/{2}s,s,\tau)
\rho(E+s,\tau)=1.\label{normal2}
\end{equation}
Similarly, the integral (\ref{Idef1}) takes the form
\begin{equation}
I(E,\tau)=-{\cal P}\!\int_{-\infty}^{\infty}\!ds\,
Y(E+\case{1}/{2}s,s,\tau)\rho(E+s,\tau)s^{-1}.\label{Idef2}
\end{equation}
By substituting the Taylor expansions
\begin{eqnarray}
&&Y(E+\case{1}/{2}s,s,\tau)=Y(E,s,\tau)+\case{1}{2}s
\frac{\partial}{\partial E}Y(E,s,\tau)+\cdots,\label{Taylor1}\\
&&\rho(E+s,\tau)=
\rho(E,\tau)+s\frac{\partial }{\partial E}\rho(E,\tau)
+\cdots,\label{Taylor2}
\end{eqnarray}
into Eq.\ (\ref{Idef2}), we obtain an expansion of $I(E,\tau)$ in higher and
higher moments $Y_{p}(E,\tau)$ of $Y(E,s,\tau)$ with respect to $s$,
\begin{equation}
Y_{p}(E,\tau)=\int_{-\infty}^{\infty}\!ds\,Y(E,s,\tau)s^{p}.
\label{momentdef}
\end{equation}
Because of the symmetry $Y(t,s,\tau)=Y(t,-s,\tau)$ only even moments contribute
[$Y_{p}(E,\tau)\equiv 0$ for $p$ odd]. Following Dyson,\cite{Dys62} we {\em
neglect the second and higher moments}. An order of magnitude estimate suggests
that the error involved in neglecting $Y_{p}$ for $p\geq 2$ is of order
$N^{-2}$.
Dyson argues that the error is actually of order $N^{-2}\ln N$, by comparison
with exact results for the distribution of the spacing of eigenvalues.

Since $Y_{-1}$ and $Y_{1}$ are identically zero, only $Y_{0}$ contributes to
$I(E,\tau)$ to second order. Substitution of the Taylor expansions
(\ref{Taylor1}) and (\ref{Taylor2}) into Eq.\ (\ref{Idef2}) yields
\begin{equation}
I(E,\tau)=-\case{1}{2}\rho(E,\tau)\frac{\partial }
{\partial E}Y_{0}(E,\tau)-
Y_{0}(E,\tau)\frac{\partial }{\partial E}\rho(E,\tau).
\label{Iapp1}
\end{equation}
Similarly, substitution of the Taylor expansions into Eq.\ (\ref{normal2})
yields
\begin{equation}
\rho(E,\tau) Y_{0}(E,\tau)=1.\label{normalapp}
\end{equation}
Combining Eqs.\ (\ref{Iapp1}) and (\ref{normalapp}) we find
\begin{equation}
I(E,\tau)=-\case{1}{2}\frac{\partial }
{\partial E}\ln\rho(E,\tau).\label{Iapp2}
\end{equation}
Hence Eq.\ (\ref{int3}) takes the form
\begin{equation}
\gamma\frac{\partial}{\partial\tau}\rho(E,\tau)=
\frac{\partial}{\partial E}\left[\rho(E,\tau)\frac{\partial }
{\partial E}\left(V(E)-\int_{-\infty}^{\infty}\!dE'\rho(E',\tau)
\ln|E-E'|+\frac{2-\beta}{2\beta}\ln\rho(E,\tau)\right)\right].
\label{int4}
\end{equation}
This is Eq.\ (\ref{DRHODTAU}), except for the final term, proportional to
$(2-\beta)/2\beta$. As noted in Ref.\ \onlinecite{Dys62}, this term is of order
$\ln N$ and can be neglected relative to the other terms, which are of order
$N$. Dropping that term, we obtain Eq.\ (\ref{DRHODTAU}).

\end{document}